\documentclass[preprint2]{aastex}

\usepackage{paralist}

\shorttitle{An interaction scenario of the KPG 302}
\shortauthors{Gabbasov et al.}

\begin{document}

\title{An interaction scenario of the galaxy pair NGC 3893/96 (KPG 302). A single passage?}

\author{R. F. Gabbasov}
\affil{Instituto de Astronom\'{i}a, Universidad Nacional Aut\'onoma de Mexico (UNAM), 
A.P. 70-264,04510, M\'exico D.F.}
\affil{Instituto de Ciencias B\'asicas e Ingenier\'{i}a, Universidad Aut\'onoma del Estado de 
Hidalgo (UAEH), Km 4.5 Carretera Pachuca - Tulancingo, Mineral de la Reforma, 42184, Hidalgo, M\'exico}
\email{ruslan.gabb@gmail.com}

\and

\author{M. Rosado}
\affil{Instituto de Astronom\'{i}a, Universidad Nacional Aut\'onoma de Mexico (UNAM),
A.P. 70-264,04510, M\'exico D.F.}

\and

\author{J. Klapp}
\affil{Instituto Nacional de Investigaciones Nucleares, Carretera M\'exico-Toluca S/N,
La Marquesa, Ocoyoacac, 52750, Estado de M\'exico, M\'exico.}
\affil{Departamento de Matem\'aticas, Centro de Investigaci\'on y de Estudios Avanzados del IPN, A.P. 14-740, 07000 M\'exico D.F., M\'exico.
}

\begin{abstract}
Using the data obtained previously from Fabry-Perot interferometry,
 we study the orbital characteristics of the interacting pair of
 galaxies KPG 302 with the aim to estimate
 a possible interaction history, the conditions necessary for the
 spiral arms formation and initial satellite mass. We found by performing N-body/SPH
 simulations of the interaction that a single passage can produce a grand 
 design spiral pattern in less than $1$ Gyr.
  Although we reproduce most of the features with the single passage, 
  the required satellite to host mass ratio should be $\sim 1:5$,
 which is not confirmed by the dynamical mass estimate made from the measured rotation curve.
 We conclude that a more realistic interaction scenario would require several
 passages in order to explain the mass ratio discrepancy.
\end{abstract}


\section{Introduction}
Galaxy interactions and mergers are considered the main morphology transformation mechanism in 
the hierarchical galaxy formation paradigm. Numerous simulations of cosmological structure formation
at different scales show that large massive galaxies and clusters are formed by accretion of 
smaller ones. With the advent of computer power and development of new methods and codes 
numerical simulations rapidly became an essential tool in theoretical study of galaxies.

First models of interacting galaxies had the aim of reproducing the spiral structure and 
extended puzzling features such as bridges and tails. Later on, more complex and detailed models 
of interacting galaxies were used to explore the history of their dynamical evolution and possible
 scenario for their future.
Numerical simulations of such systems are very difficult and challenging due to a large 
space of parameters that should be covered before an acceptable solution is found.
For this purpose, two methods are used: a backward orbit
 integration from matched current configuration and fitting the evolved positions from 
 some initial ones guessed by the ``simulator intuition''.
Although the problem is in general, degenerate and may have multiple solutions, the
determination of at least one already gives important information on dynamical evolution 
and can be confronted with the observations. 

The parameter space may be reduced to some degree making several assumptions, such as a rigid halo, 
point mass distribution, unique passage, etc.
Nevertheless, even with such substantial assumptions the CPU-time needed for detailed
self-consistent simulations is unacceptable.
Genetic algorithms were successfully applied for rapid automatic exploration of orbital parameters
of several interacting pairs e.g., \citet{Wahde1998,Theis2001}. These models were studied using
 restricted N-body method that ignores self gravity of particles and reduces the problem to
 N single body problems.
The method does not necessarily reproduce internal morphological features such as 
the spiral structure or a bar that may be obtained with self-consistent simulations.
However, the most important results of such simulations come from the galaxy mass distribution, 
which influences the tidal features of interacting systems and allows for orbital decay.
In order to take into account the effects of extended mass distribution dynamical friction was 
included in MINGA code for orbit parameters search developed by \citet{Petsch2008}.
A new tool called \verb+Identikid+ that combines test-particle technique and self-consistent simulations
was recently developed \citep{Barnes2009}. These authors were successful in reproducing dozens 
of interacting pairs and made the code publicly available.

Minor mergers such as the M\,51 pair with masses of galaxies differing several times
are very interesting since they allow us to study induced spiral arms and bar formation, 
dynamical disk heating, and properties of dark halo mass distribution.
The first numerical simulation of a minor merger is described in the seminal paper of \citet{TT1972}, 
where they successfully reproduce the grand design spiral pattern of M\,51 system.
Since then more sophisticated models have appeared and the M\,51 became a standard interacting 
system to study in numerical simulations of galaxies.
\citet{Salo2000a} and \citet{Salo2000b} performed a very thorough study and obtained the
orbital parameters that reproduce the morphological and kinematic features in scenarios of 
single and multiple passages.
A rigid halo used in their models does not enable the mass loss, dynamical friction and angular momentum redistribution due to the interaction process. They explore these limitations using models with a
dynamical friction approximation and also with a live halo and conclude that the latter leads to a faster orbital decay and less disk heating.

Recently, a high resolution model of M\,51 was reproduced by \citet{Dobbs2010} where a comparison of the pattern speed and the pitch angle of spiral arms with observations was made. The halo still was 
represented by static potential and the companion was represented as a point mass.

In this work we study the interaction process between two galaxies forming an isolated pair KPG 302 with the aim of determining the orbital configuration, the origin of the satellite, and the 
possibility of producing the grand design spiral pattern by interaction.
Using available observational data we construct the galaxy models and search for the orbit
configuration. We describe the orbital parameters that reproduce the observed morphology and 
kinematics of the system, and discuss the results and limitations of our models.

\section{Observational data}
\begin{figure}[!htp]
\plotone{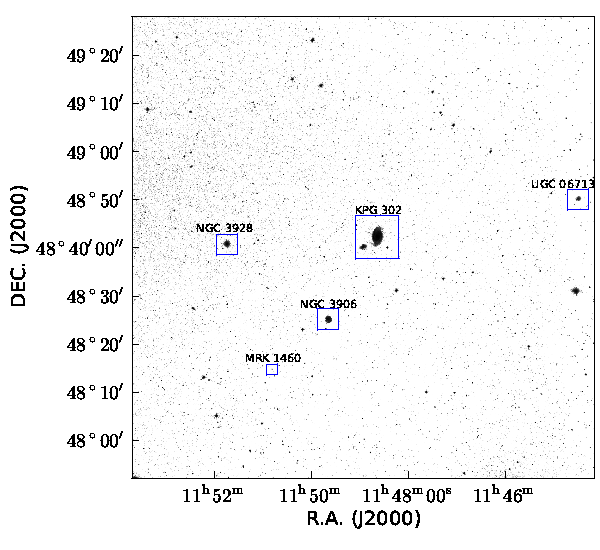}
\figcaption{The $1.6\degr\times1.6\degr$ DSS field around KPG 302. Bright neighbor galaxies are indicated.
\label{dss-environment}}
\end{figure}

\begin{figure}[!htp]
\plotone{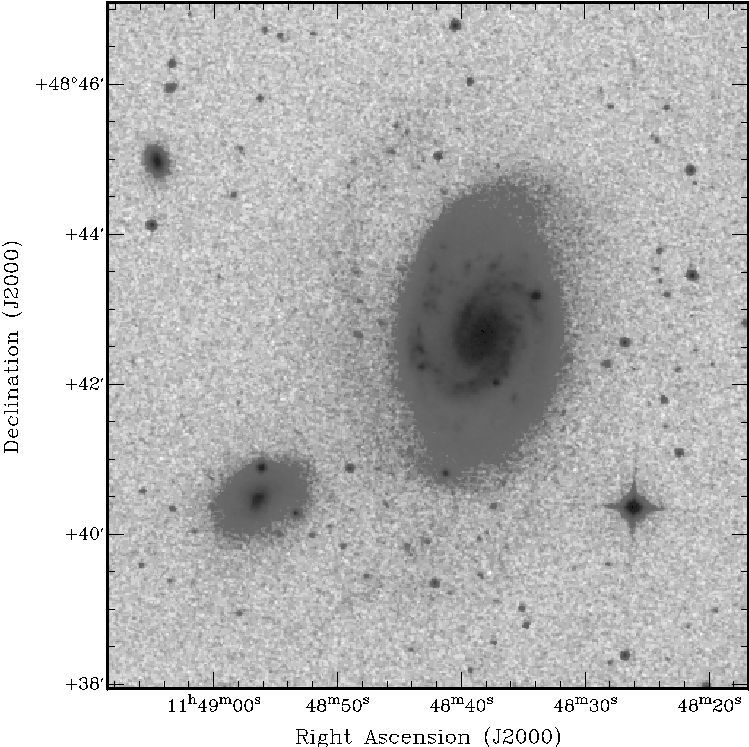}
\figcaption{DSS2 $R$ band image of KPG 302. Contrast is enhanced to depict the tidal features observed
as a faint spiral arm on the east side of the host and a dim bridge between the galaxies on the south.
\label{dss2r}}
\end{figure}

KPG 302 is an isolated galaxy pair composed of a grand design spiral galaxy NGC 3893 (host) and 
a companion dwarf galaxy NGC 3896 (satellite), located in the Ursa Major cluster.
The environment of the pair is illustrated in Fig.~\ref{dss-environment}, which shows the DSS2 
$R$ image of $1.6\degr\times1.6\degr$ field centered at NGC 3896. The neighbors brighter than
$17$ mag and having systemic velocity difference $\pm 500$ km s$^{-1}$ are marked. 
The closest neighbor NGC 3906 is lies $20\arcmin$ south-east from NGC 3893.
The tidal features can be observed from high contrast optical images, such as $R$-band DSS2 image in Fig.~\ref{dss2r}, UV images, and in HI continuum image. A faint spiral arm extends from the south
 to north side of the host making an arc on the east side. Stellar debris can also be observed on the
 south side tracing a bridge to the satellite. This bridge is better seen in $B$ band images, 
suggesting a younger stellar population.

Information about KPG 302 properties is spread among papers devoted to statistical studies,
or large surveys mainly of the host galaxy.
\citet{Meyssonnier1983} studied kinematics of the host and obtained the optical rotation curve 
and the dynamical mass. The discrepancies in velocity field and excitation rates were suggested to 
be due to proximity with the satellite.
Later on, the study in HI line revealed a large envelope of hydrogen that encompasses both galaxies
\citet{Verheijen20014,Verheijen20015}.
They showed that the HI disk of NGC 3893 changes inclination and 
Position Angle (PA) starting from a radius of $1.4\arcmin$ ($\sim 5.5$ kpc) which also coincides 
with the decay of the HI Rotation Curve (RC). These effects indicate that the system is 
involved in an ongoing interaction process.

\begin{figure}[!htp]
\plotone{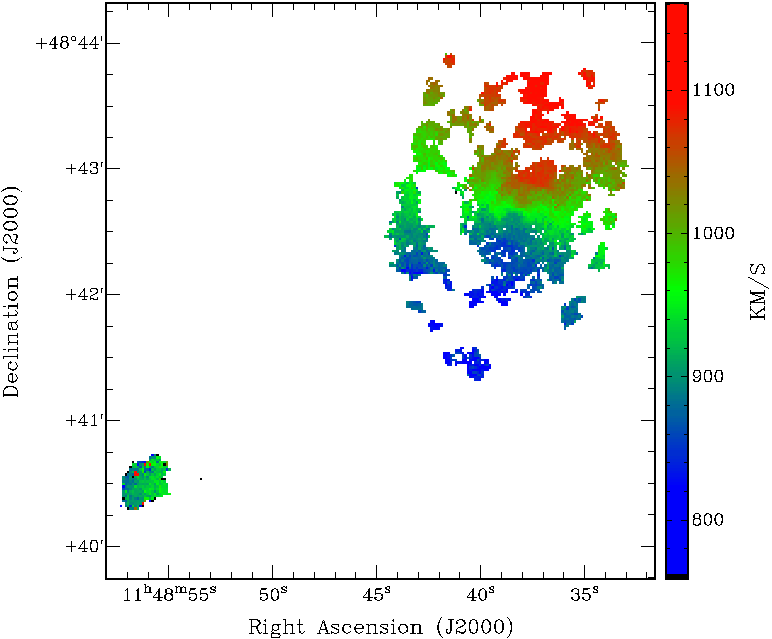}
\figcaption{Velocity map (first moment of H$\alpha$ cube) of KPG 302 \citep{Fuentes2007}.
\label{obs_mom}}
\end{figure}

A recent study of the KPG 302 kinematics using Fabry-Perot 
H$\alpha$ interferometry \citep{Fuentes2007} (hereafter Paper I) has shown the presence of many 
regions with non-circular motions in the south part of the disk of NGC 3893 which is close to 
the companion NGC 3896. The companion shows an enhanced star formation, probably caused by
interaction. The first moment map from their study is shown in Fig.~\ref{obs_mom} and the main observed
properties of the galaxies are summarized in the Table \ref{tab1}.

\begin{table}
 \centering
 \begin{minipage}{140mm}
  \caption{Main observed parameters of the KPG302 members.}
  \begin{tabular}{@{}lll@{}}
  \hline
  & NGC 3893 & NGC 3896\\
\hline
 Coordinates & $\alpha=11h 48m 38.38$  & $\alpha=11h 48m 56.39s$ \\
 (J2000) & $\delta=+48\degr 42\arcmin 34.4\arcsec$  & $\delta=+48\degr 40\arcmin 28.8\arcsec$ \\
 Morph. type & SABc & SB0/a\\
 V$_{max}$\tablenotemark{a}, km s$^{-1}$ & $197\pm 10$ & $50\pm 10$ \\
 P.A.\tablenotemark{a}, \degr & $340\pm 10$ & $294\pm 5$ \\
 $i$\tablenotemark{a}, \degr & $45\pm 3$ & $49\pm 3$ \\
 V$_{sys}$\tablenotemark{a}, km s$^{-1}$ & $967\pm 5$ & $920\pm 5$ \\
 D$_{25}$\tablenotemark{b}, $\arcmin$ & $4.5$ & $1.4$ \\
 h$_R$, $\arcsec$ & 24\tablenotemark{c} & 18 \tablenotemark{c}\\
   & 28\tablenotemark{d} & --\\
\end{tabular}
\tablenotetext{a}{\citet{Fuentes2007}}
\tablenotetext{b}{From NED}
\tablenotetext{c}{\citet{Laurikainen2000}}
\tablenotetext{d}{\citet{Kassin2006a}}
\end{minipage}
\label{tab1}
\end{table}

\citet{Kranz2003a} argued, on the basis of young stellar population
 that the interaction had taken place on a short time scale.
They also determined the corotation radius and pattern speed by matching 
the $K'$-band spiral pattern with 2D hydrodynamical simulations. 
They found that spiral arms within $65\arcsec$ fit very well with a maximal disk
model and gave a disk mass $M_d=2.32\times 10^{10}$ M$_{\odot}$. The corresponding
corotation radius is $R_{CR}=5.5\pm 0.5$ kpc and the pattern speed is
$\Omega_{P}\approx 38$ km s$^{-1}$ kpc$^{-1}$ (see also \citet{Slyz2003}).

Detailed $BVRI$ photometry \citep{Laurikainen1998, Laurikainen2000}
of this pair permitted to perform the bulge-disk decomposition (see Table~\ref{tab2}),
and the color difference measurement.
Recently, \citet{Kassin2006a} studied the optical surface photometry 
of the NGC 3893 and found that the disk can be fitted with an exponential profile with a
scale length $h=28\arcsec$ and the central surface brightness $\mu_0=16.9$
mag $/(\arcsec)^{2}$. They found from the best fit of the
bulge and disc profiles a rather massive bulge component B/D$=0.29$.

On the other hand, recent estimations by \citet{Torres2011}
give the mass of the stellar and gas disk $M_{*}\ga 2.6\times 10^{10}$ M$_{\sun}$ and 
$M_{gas}\approx 4.1\times 10^{9}$ M$_{\odot}$, respectively.

With the above data it is possible to decompose the RC and calculate the dark 
halo mass distribution. \citet{Fuentes2007} have fitted several
halo density profiles to the RC, and found that it was
difficult to reduce $\chi^2$ errors for all the halo profiles they used.
Here we redo the fitting obtaining new parameters for dark halo component.

\section{Galaxy Models}
For host galaxy models we adopt the baryonic mass $M=2.28\times 10^{10}$ M$_{\sun}$ 
and complement the missing mass with the Hernquist dark matter halo profile \citep{Hernquist1990}
in order to match the observed RC. We choose the Herquist density profile which mimics the cuspy 
NFW density profile except the difference at large radius is not important in process 
of tidal tail formation \citep{Dubinski1999}. The resulting ratios of baryonic to dark matter 
of the models are rather large, particularly for the satellite ($\sim 30$) model.

We construct our models following the procedure described by \citet{Hernquist1993}.
Details of the models are described in \citet{Gabbasov2006}. Both stellar and gas masses are
distributed in exponential disks embedded in the Hernquist halo.
The stellar disks have minimum value of the Toomre stability parameter Q$_{star}=2.0$.
The gas component is distributed with the same scale radius 
as the stellar disk but with 10 times smaller scale height.
We assume the gas component obeys the isothermal equation of state with a  
temperature $10^4$ K, i.e., cooling and heating processes cancel each other 
exactly and the energy due to shocks rapidly radiates \citep{vdk1984}. 
Since we are also interested in the mass transfer into the center of the host by interaction,
and possible bulge formation, we do not put significant mass into the bulge.
A small bulge is added only to the host galaxy in order to prevent  possible 
tidally induced bar formation (see Table~\ref{tab2}). The initial conditions of the satellite are
rather uncertain, and we adopt the same model but scaled to have the total mass determined
by procedure described in the next section.

\begin{figure}[!htp]
\includegraphics[width=\columnwidth]{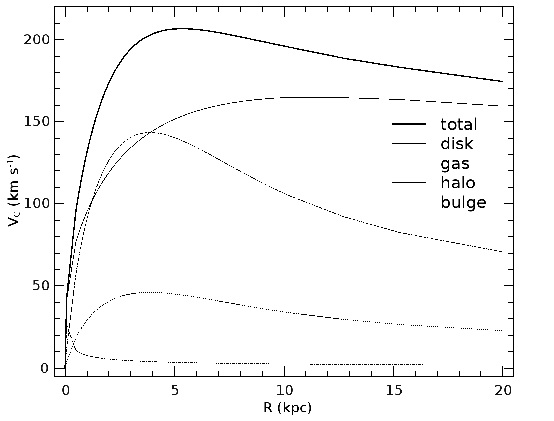}
\includegraphics[width=\columnwidth]{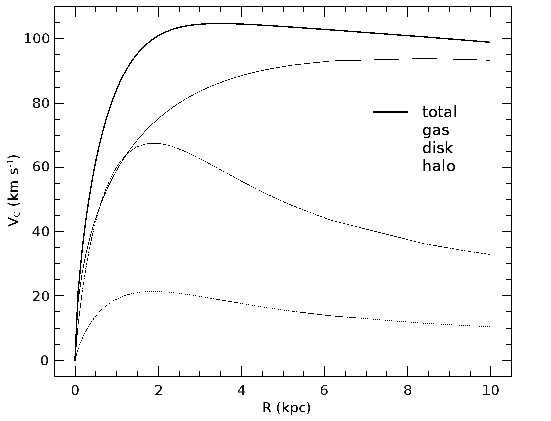}
\figcaption{Results of the decomposition ofthe  circular velocity of the host (top) 
and the satellite (bottom).
\label{rc-decomp}}
\end{figure}

The initial RCs of the host and satellite galaxy models are shown in Fig.~\ref{rc-decomp}.
 Note that the maximum velocity for the satellite is twice the measured value
(see \ref{tab1}). Since we do not know the initial RC we expect that tidal interaction will
disrupt the satellite and reduce the amplitude of the RC to the observed one. The satellite to host mass
 ratio is $\sim 0.2$.

Our fiducial model consists of
 $2.3\times 10^5$ particles for the stellar disk, $2.7\times 10^4$ particles for the gas disk, 
 $6.4\times 10^3$ particles for the bulge, and $4.2\times 10^6$ particles for the halo.
The satellite is represented as a scaled down model of the host, totaling $8.5\times 10^5$ particles.

We use the Gadget 2 TREE/SPH code \citep{Springel2005} with adaptive timesteps, force tolerance
parameter $\Theta=0.5$ and softening length $\varepsilon = 0.0015$ (60 pc) for all particles. 
The adopted value for $\Theta$ is rather large, which might influence the precision of
gravitational force calculation and would be reflected in poor linear momentum conservation.
Our simulations conserve the momentum within $4\%$ which we consider satisfactory.

Our tests varying the number of particles,
their ratio and $\varepsilon$ confirmed our previous findings \citep{Gabbasov2006} 
and the results described by \citet{Farouki94}, namely, that unequal
mass particles lead to excessive disk heating and models are prone to bar formation.
Despite the higher cost paid for increased halo resolution the models are more stable and lead 
to richer tidal structures being produced in collision simulations below.

Since the initial conditions are not in exact equilibrium, a relaxation period is necessary
during which small concentric undulations in the disk and readjustment is observed in the RC shape.
We evolved both galaxies in isolation in order to ensure both are stable against bar
formation. We found no evidence of bar or any strong non-axisymmetric instability
 after evolving the models for $2$ Gyr, with the Toomre
 stability parameters remaining nearly constant in both galaxies (see Table \ref{tab2}).
The outputs at $0.25$ Gyr are taken as relaxed galaxy models for interaction simulations.

We also explored host models evolved with a ``rigid'' halo potential instead of a ``live'' halo.
In these models the gravitational potential of halo particle distribution was replaced by the
 corresponding static potential. We evolved the host in isolation for $4$ Gyr
in order to investigate the development of any disk instability. The system rapidly relaxed 
to a smooth axisymmetric state without any significant disk heating
($Q_{star}=Q_{gas}=2.0$), or change in the shape of the RC.

\begin{table}
 \centering
  \caption{Galaxy model parameters.}
  \begin{tabular}{@{}ccc@{}}
  \hline
 Parameter & NGC 3893 & NGC 3896\\
\hline
 M$_b$, $10^{10}\, \mathrm{M}_\odot$ &  0.055  & --\\
 M$_d$, $10^{10}\, \mathrm{M}_\odot$ & 2.0 & 0.22 \\
 M$_g$, $10^{10}\, \mathrm{M}_\odot$ & 0.23 & 0.022 \\
 M$_h$, $10^{10}\, \mathrm{M}_\odot$ & 35.2 & 6.6 \\
 a$_b$, kpc & 0.32 & --\\
 h$_g$, kpc & 1.8 & 0.8\\
 h$_d$, kpc & 1.8 & 0.8\\
 a$_h$, kpc & 12 & 8\\
 z$_d$, kpc & 0.28 & 0.28\\
 z$_g$, kpc & 0.028 & 0.028\\
 Q$_{star}$ & 2.1   &  2.2 \\
 Q$_{gas}$  & 2.1   &  2.2 \\
\end{tabular}
\label{tab2}
\end{table}

\section{Orbital parameters search}
According to works of \citet{Benson2005} and \citet{Khochfar2006}, 
the parabolic orbits are the dominant ones at the early stage of the galaxy interactions.
 The hyperbolic orbits have much smaller probability and have large velocities that would not
 lead to merging, whereas elliptic orbits rather represent late stages of the encounter and 
 lead to a rapid merging.

Previous numerical simulations have shown that the strength of perturbation
depends on many different factors, including the galaxy mass ratio, the 
pericentric separation, the impact velocity, and also the disks orientations.
Analyzing the images and H$\alpha$ kinematics we may conclude that the interaction process 
barely affects the inner part of the disk of the host galaxy, but rather seems to be responsible for 
creation of outer spiral arms. This could be explained by a rather weak tidal perturbation 
due to satellite, but, on the other hand, the perturbation waves may have ceased during the 
time of the last close approach.
Here we assume that the tidal forces are less efficient in the inner part of the 
host's disk which remains practically unperturbed.

\begin{figure}[!htp]
\plotone{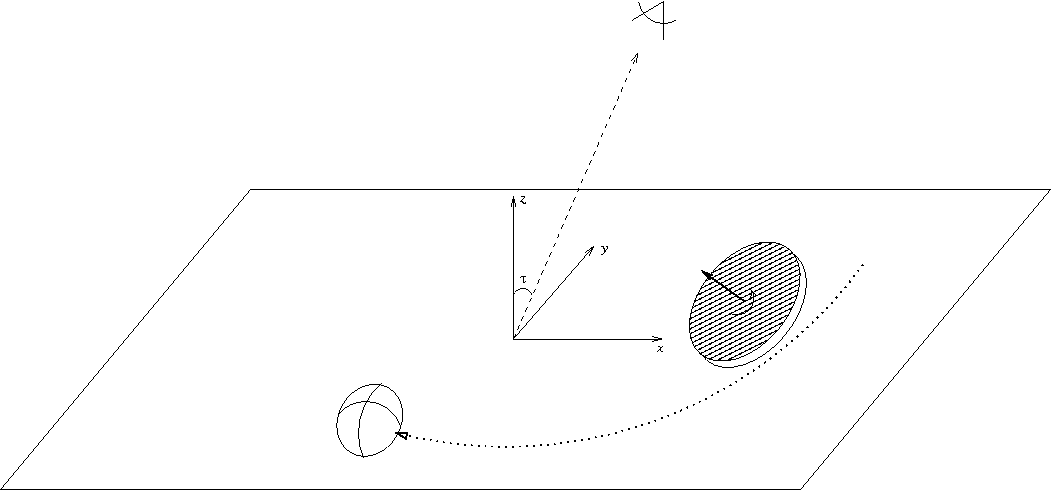}
\figcaption{Sketch of the orbit plane, coordinate system and the observer location.
\label{fig3}}
\end{figure}

To describe the two-body parametric orbit one needs to know the pericentric separation, $p$,
eccentricity $e$, and the time of pericenter $t_p$. We define the time of pericenter as 
the time needed to reach the point of the closest approach of point masses from a given
initial position. The orientation of the disk is given by the disk inclinations $i$, 
and the position angle (PA). 
The framework used for the orbital parameter search is shown in Fig.~\ref{fig3}. Here the $y$ 
axis corresponds to the north and the $x$ axis to the west. 
An observer is located at an angle $\tau$ from the $z$ axis along the $x=const$ plane.
Moreover, the line of nodes of the orbit (i.e., the line between the focus and the origin of the
 coordinates), is rotated by some angle $\phi$ along the $z$ axis.

In order to reduce the orbital parameter space, we made the following assumptions:
\begin{inparaenum}[\itshape i\upshape)]
\item the encounter is parabolic, 
\item the initial mass ratio of the galaxies is fixed, 
\item the satellite just made a first fly-by passage, 
\item the spiral pattern was induced by the companion due to tidal interaction, and 
\item the spiral arms of the host galaxy are trailing.
\end{inparaenum}

As a first step we have fixed the view angle $\tau=0\degr$, and set PA and
$i$ of the host galaxy to match the observed ones.
For orbital parameters search we simply model the satellite as a Hernquist sphere whose 
parameters are total mass and scale radius are set to $M_h = 6.84\times10^{10}$ M$_\odot$
and $a_h = 8$ kpc, respectively.
First, we perform runs with a small number of particles ($N=3\times 10^5$). 
The satellite is launched on the parabolic orbit ($e=1.0$) with $p=5.6$ kpc and
$t_p=175$ Myr. In each run the satellite orbit is rotated successively along the $z$-axis with 
a step $\phi=90\degr$. Then we change the view point to the opposite one ($\tau=180\degr$), 
and also change the direction of rotation of the host galaxy, and repeat the runs.
Thus, we have performed eight runs that cover both prograde and retrograde orbits, 
giving us the clues of a possible single or double passage orbit. The aim of these runs is to 
find configurations in which the satellite is found in the third quadrant of the reference frame.
In polar coordinates centered at the host galaxy the separation should be $\rho\approx 3\arcmin.7$
($\sim 15$) kpc and polar angle $\varphi\approx 220\degr$ from the positive $x$ axis.

A correctly determined orbit must reproduce the projected position of the satellite and its radial
 component of the peculiar velocity. In all eight runs the peculiar velocity of the satellite was 
 close to zero. Thus, next we adjust $\tau$ in order to obtain a projected velocity difference 
 close to the observed one ($\sim 40$ km s$^{-1}$) and obtain $\tau\approx 25\degr$.
 We continue to restrict these runs further by now changing the angle $\phi$ by $10\degr$ within
 the sector of $45\degr$ centered at the angle $\phi=90\degr$ found from our first iteration runs.

Finally, the newly determined orbital parameters are used to setup a full satellite model 
and next series of simulations. The last two unknowns to be determined are the  inclination and 
PA of the satellite. We tried combinations for inclination of $0\degr$, $\pm 30\degr$ and PA$=0\degr$, $\pm 90\degr$.
The best configuration is then repeated with high resolution.
The fiducial parameters that match the comparison criteria for a single passage are 
listed in Table~\ref{tab3}. These correspond to the satellite falling from the north with $\tau=25\degr$ and $\varphi=90\degr$ 
on the retrograde parabolic orbit.
Both galaxies appear to be rotating counterclockwise from the observer's point of view.
 
\begin{table}
 \centering
  \caption{Initial positions and velocities of members of KPG 302 in the fiducial model.}
  \begin{tabular}{@{}cccc@{}}
  \hline
 Parameter & & NGC 3896 & NGC 3893\\
\hline
 Position, kpc  & $x$ &  5.3 & -0.9 \\
 				& $y$ & 55.1 & -10.0 \\
 				& $z$ &  3.0 & -0.5
\medskip\\
 Velocity, km s$^{-1}$  & $v_x$ &    9.7 & -1.7 \\
		 				& $v_y$ & -202.9 & 37.0 \\
	 					& $v_z$ &    5.6 & -1.0
\medskip\\
 P.A., $\degr$ & & 0 & 175 \\
  $i$, $\degr$ & & 30 & 45 \\
\hline
   \end{tabular}
\label{tab3}
\end{table}

\section{Results}
\begin{figure}[!htp]
\plotone{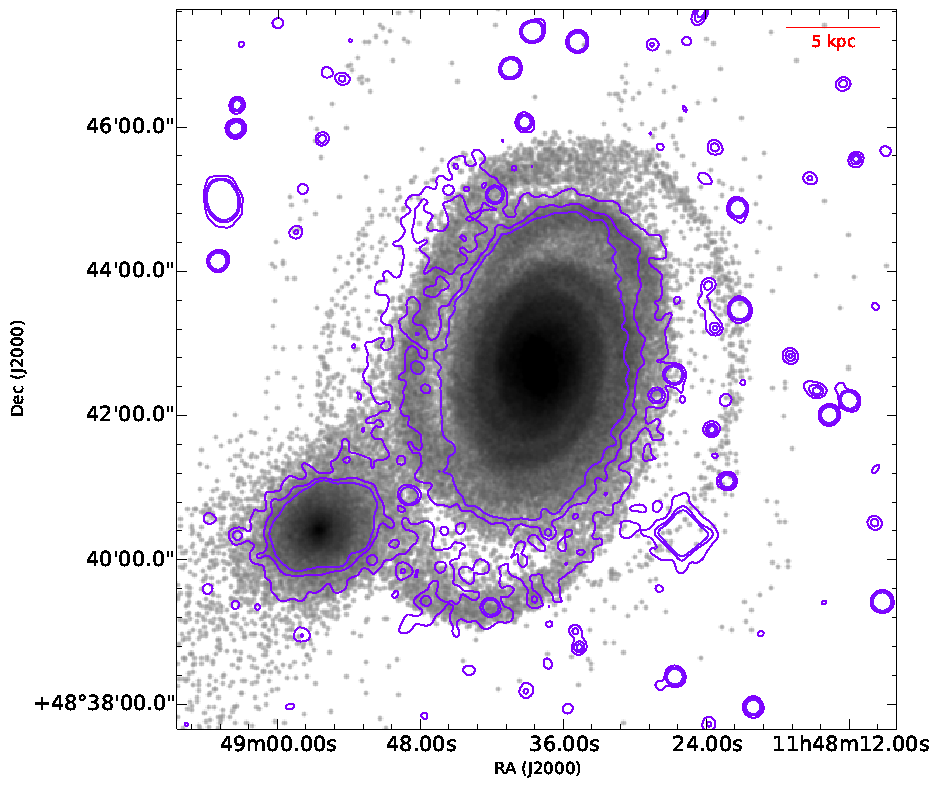}
\figcaption{Logarithmic surface density of the stellar disk particles for the snapshot at $t_{obs}=0.97$ Gyr. The 
contours show the faint features extracted from the Fig.~\ref{dss2r} at 
levels of $1\%$, $3\%$ and $6\%$ of the background mean.
\label{snap}}
\end{figure}

After performing series of simulations we have determined the parameters that roughly reproduce the
observed galaxy pair. The time elapsed from the start of the simulation and the optimal configuration
that matches the observations is $t_{obs}=0.97$ Gyr. The corresponding snapshot is show in 
Fig.~\ref{snap} as logarithm scaled projected surface density.
The Table~\ref{tab4} lists the final positions and velocities of the galaxies
 which were measured by locating iteratively the centers of mass of each galaxy.
 This procedure leads to rather large errors, and we estimate the values listed in 
 the table to have relative errors $\leq 10\%$.
 As the Table~\ref{tab4} illustrates, along the line of sight the satellite is $\sim 4$ kpc closer than the host, 
 and it moves towards the observer appearing blueshifted in velocity maps.

We obtain a projected separation of $16.0$ kpc and a line of sight velocity difference of 
$63$ km s$^{-1}$. The latter is somewhat larger than the maximum systemic velocity difference found
 in Paper I, but is in excellent agreement with that given by the NED database,
$62\pm 36$ km s$^{-1}$ with the error due to uncertainty in satellite recession velocity.
We neglect a small $1$ kpc difference in the projected separation since it is determined by the
instant position of the satellite, which may be further reduced with fine tuned simulations.

\begin{table}
 \centering
  \caption{Final positions and velocities of members of KPG 302 at $t=0.97$ Gyr.}
  \begin{tabular}{@{}cccc@{}}
  \hline
 Parameter & & NGC 3896 & NGC 3893 \\
\hline
  Position, kpc & $x$ &  -5.2  & 4.9 \\
                & $y$ &   0.0 & 12.5 \\
                & $z$ &  -1.7 & 2.0
\medskip\\
 Velocity, km s$^{-1}$  & $v_x$ &  187.1 & -0.3 \\
			 			& $v_y$ &  212.5 &  3.7 \\
			   			& $v_z$ &   63.4 &  0.6
\medskip\\
 P.A.\tablenotemark{a}, $\degr$ & & 145 & 168 \\
				   $i$, $\degr$ & & 41 & 45 \\
\hline
   \end{tabular}
\tablenotetext{a}{Kinematic PA measured for stellar disc.}
\label{tab4}
\end{table}

Once a plausible orbital configuration is found, we proceed with the detailed comparison.
As a first step we make a morphological comparison of tidal features and then proceed 
with an analysis of the kinematic data, such as the velocity field and the RC.

\begin{figure}[!htp]
\plotone{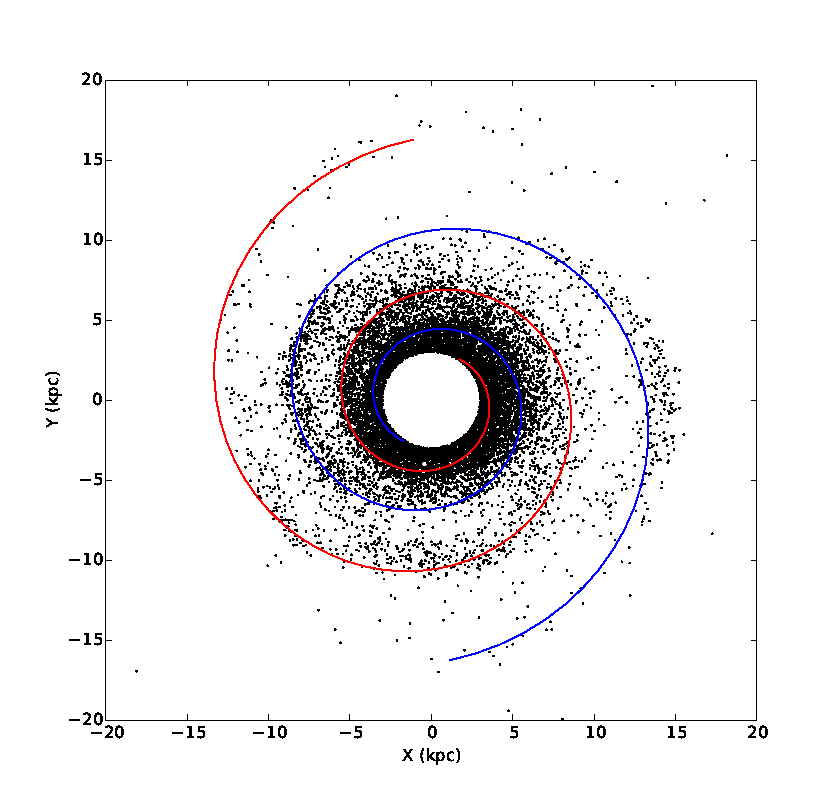}
\figcaption{Deprojected stellar disk of the host with overplotted logarithmic spirals $\theta=\kappa\log (R/4.4)$ with pitch angle $\kappa=8\degr$. Only every tenth particle is plotted, and the region $r<3$ kpc is excluded in order to avoid particle crowding.
\label{sparm}}
\end{figure}

Tidally induced spiral arms of the host can be quantified by measuring their pitch angles. 
Figure \ref{sparm} shows the deprojected disk particles in order to depict the shape of the spiral 
arms and their degree of windup. The spiral arms in stellar and gas disks are quite symmetric and 
have pitch angle $\kappa=8\degr$. This is much smaller than the value of $19\degr$ found from
a deprojected image of NGC 3893 \citep{Kranz2003a}.
We also estimated the spiral pattern speed to be in the range $\Omega_P=24$--$36$ km s$^{-1}$ kpc$^{-1}$ 
by measuring the pattern displacement in a thin cylindrical shell at radius $3$ kpc.
This gives the respective corotation radius range $R_{CR} = 8$--$6.5$ kpc.
\begin{figure}[!htp]
\plotone{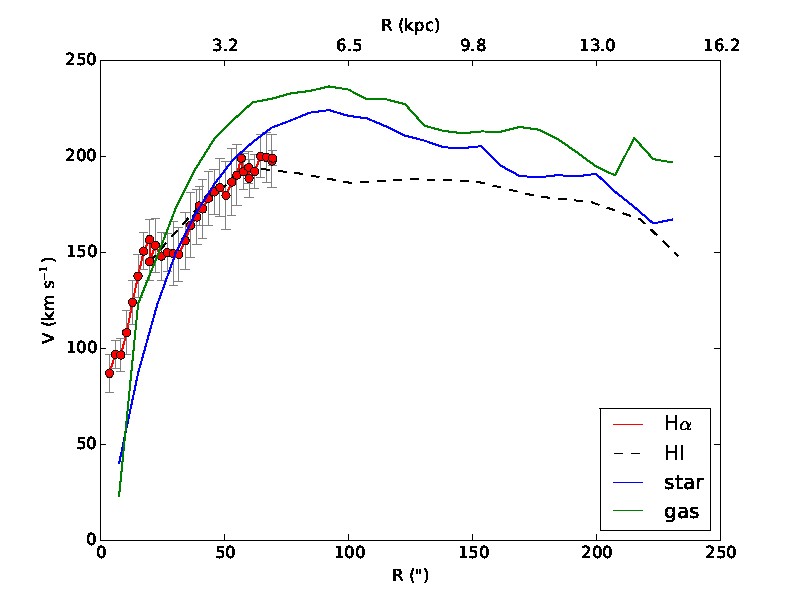}
\figcaption{Final rotation curve of the stellar (blue) and gas (green) disk of the host galaxy
as compared to the observed H$\alpha$ (red dots \citet{Fuentes2007}) and HI (dashed \citet{Verheijen20014}) rotation curves.
\label{obs-rc}}
\end{figure}

\begin{figure}[!htp]
\plotone{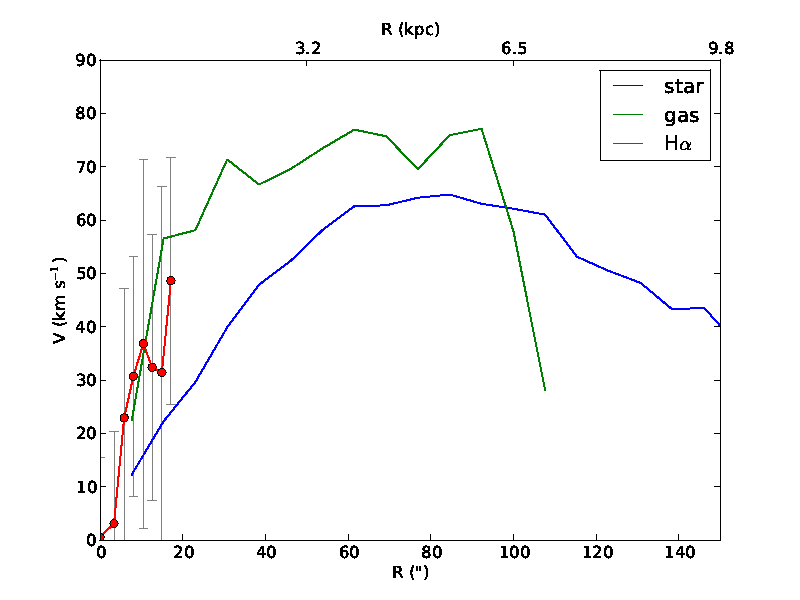}
\figcaption{Final rotation curve of the stellar (blue) and gas (green) disk of the 
satellite galaxy as compared to the observed H$\alpha$ (red dots \citet{Fuentes2007}).
\label{rcsat}}
\end{figure}

\begin{figure}[htp]
\includegraphics[width=\columnwidth]{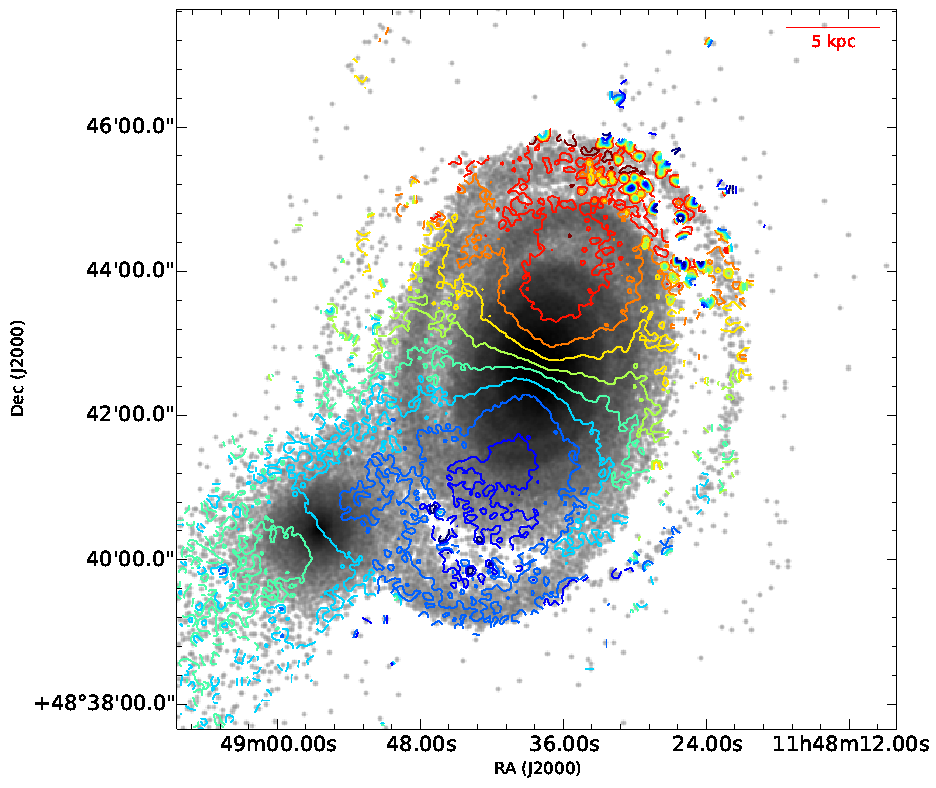}
\includegraphics[width=\columnwidth]{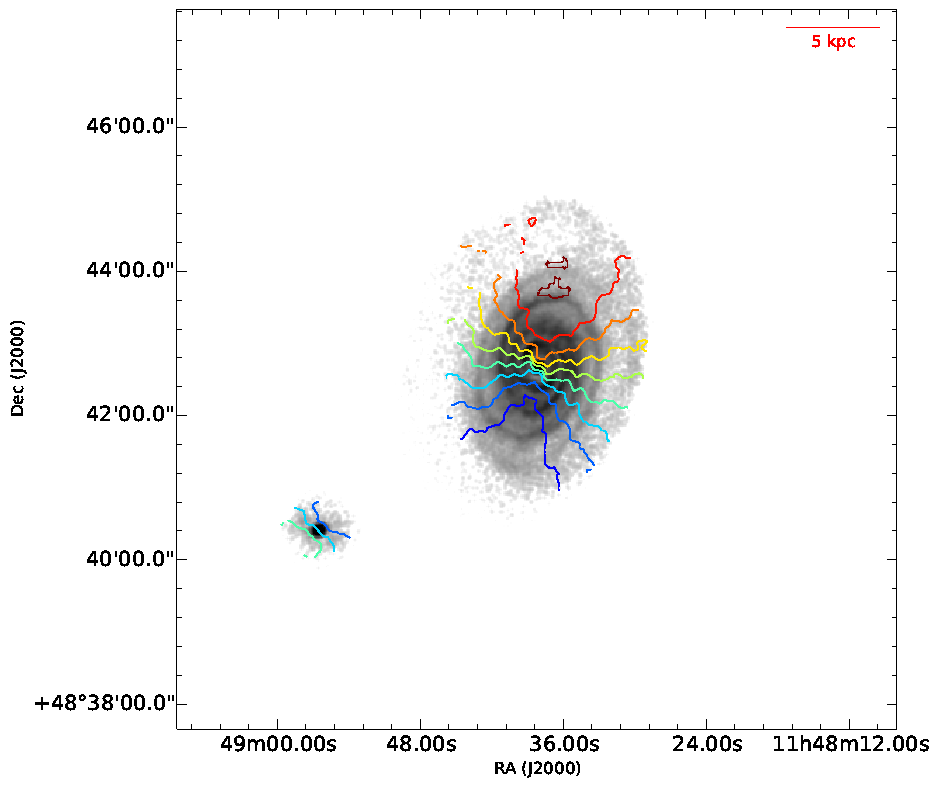}
\figcaption{Zeroth moment maps of the stellar (top) and gas (bottom) components with the
 corresponding velocity contours. The logarithmic scale is applied to the intensities, and the contour levels are separated by $50$ km s$^{-1}$.
\label{momaps}}
\end{figure}

In order to be able to process the results with astronomical software 
we created 3D data cubes in fits with the format $512\times 512\times 48$ from the output snapshots, 
and smoothed them out with a Gaussian function of FWHM equal to the seeing of the observing 
site ($\sim 2\arcsec.5$). We further assume the scale $1\arcsec=65$ pc and $1$ pix$=1\arcsec.16$.
Then we apply the same reduction procedure used for analysis in Paper I, except that the 
continuum was not subtracted. We compare the corresponding RCs 
together with the observed curves in Figs.~\ref{obs-rc} and \ref{rcsat}. The resulting moment maps for the stellar and gas components 
are shown in Fig.~\ref{momaps}.


We conclude that the model of a single passage reproduces the overall characteristics of KPG 302.
The two main criteria used for comparison are satisfied, namely, the projected separation and the
velocity difference along the line of sight.
In order to illustrate the morphological details we overplot the contours extracted 
form DSS2 $R$ image at levels of $1\%$, $3\%$, and $6\%$ of the background level onto 
the surface density projection of the stellar disk. 
 Our model predicts the existence of a weak spiral arm on the west side and 
 faint stellar debris on the north-west side of the host, which is, however, not seen in
 Fig.~\ref{dss2r}. In addition, a very large and faint tidal tail is formed by disrupted satellite 
 stars extending $20\arcmin$ from the host in south-east direction.

The rotation curve of the host is strongly affected because of interaction for R$>150\arcsec$. 
Moreover, the maximum of the modeled RC for stellar disk is larger than observed by $\sim 20$ km s$^{-1}$.
In addition, there is a small difference between stellar and gas RC. These differences may
be due to the fact that we use a single value for $i$ and PA when deriving the RCs, which
are not necessarily constant along the radius, or due to the different rates of orbital angular 
momentum absorption by the stellar and gas components. Except for the maximum velocity difference 
between the modeled and observed values, the RC is satisfactorily reproduced.

Concerning the properties of the satellite, the first thing that is apparent from 
the previous moment maps is that the satellite does not disrupt significantly
and velocity contours are regular and symmetric.
Most of the gas of the satellite disk is now concentrated in the center. This may, in principle, lead to 
an enhanced star formation, indications of which are seen in photometric color indexes gradient
\citep{Laurikainen1998}.

Another important result is that the satellite changes its initial disk
orientation (see Table \ref{tab3}) because of interaction.
The interaction tends to align the satellite spin with that of the host, possibly from a rotating
satellite halo that gained its spin from orbital motion, which, in turn, involves the disk particles.
We consider this phenomenon to be worth further exploration.
We tried different initial inclinations and P.A.s of the satellite in order to match both the kinematic 
and photometric PAs simultaneously, but without success, probably because not all configurations
 were covered. The rotation curve of the satellite is shown in Fig.~\ref{rcsat}. The first thing to note
 is that gas and stellar components show very different amplitudes and shapes starting from the origin.
 The stellar RC decay is quite expected and shows a maximum amplitude of $\sim 63$ km s$^{-1}$
 at $5.5$ kpc. The high gas concentration in the center would explain the steeply rising gas RC.
 
From the above results it is expected that the halo properties would play a major role 
in the interaction process.
In particular, the dynamical friction will lead to the transfer of the orbital angular momentum
into the internal momentum of halos. In order to explore the possible differences in the interaction
 process and host properties we repeated our fiducial simulation replacing the host galaxy with
  the rigid halo model and keeping the satellite with live halo. 
  In this case we leave the host static, letting only the satellite to move.
 This led to some disk wobbling caused by tidal perturbation of the approaching satellite 
  which erased the internal spiral arms. It is clear that such model will not correctly follow
  the interaction process. However, we found that it is sufficient to detect main differences
  between halo models as follows.
 
\begin{figure}[!htp]
\includegraphics[width=\columnwidth]{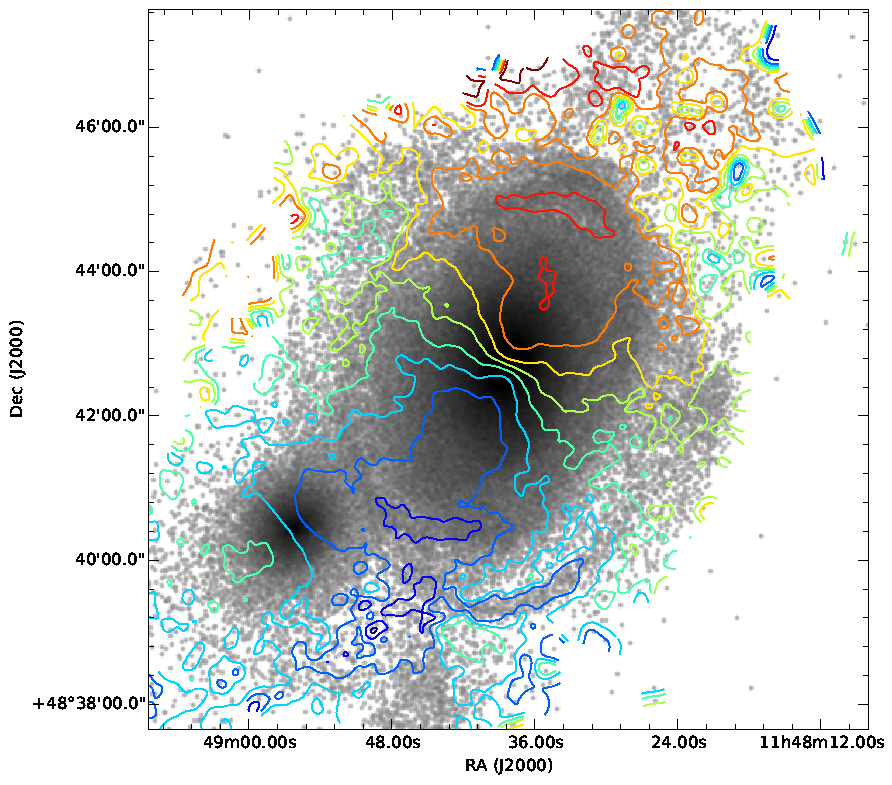}
\includegraphics[width=\columnwidth]{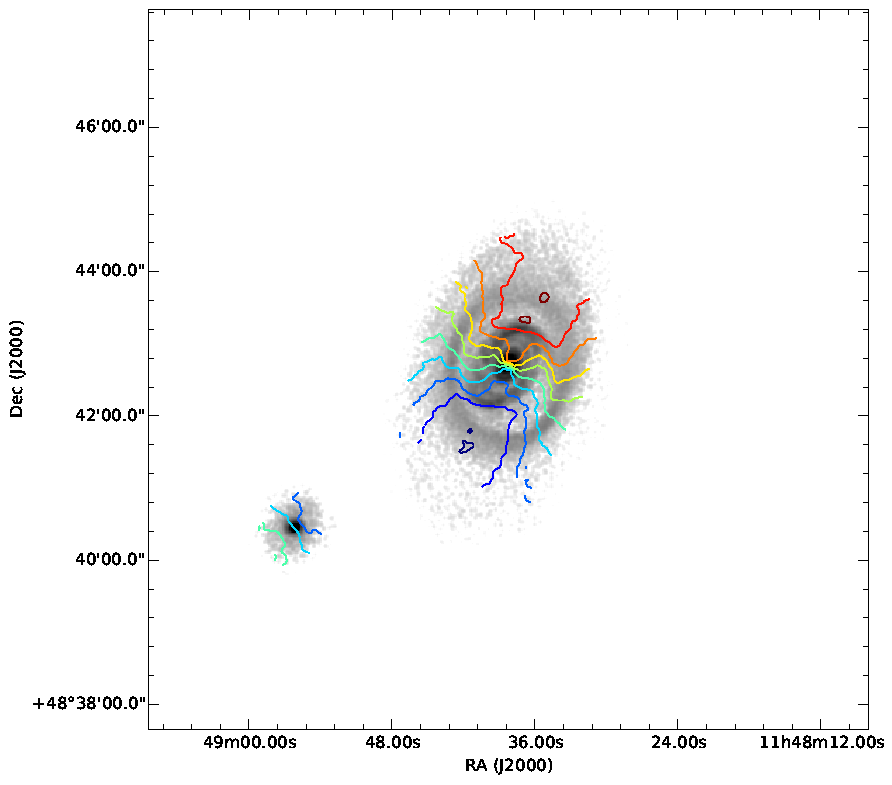}
\figcaption{Zeroth moment maps of the stellar (top) and gas (bottom) components with
 corresponding velocity contours for static host with rigid halo evolution. The scale is the same as in Fig.~\ref{momaps}
\label{momapsrig}}
\end{figure}

\begin{figure}[!htp]
\plotone{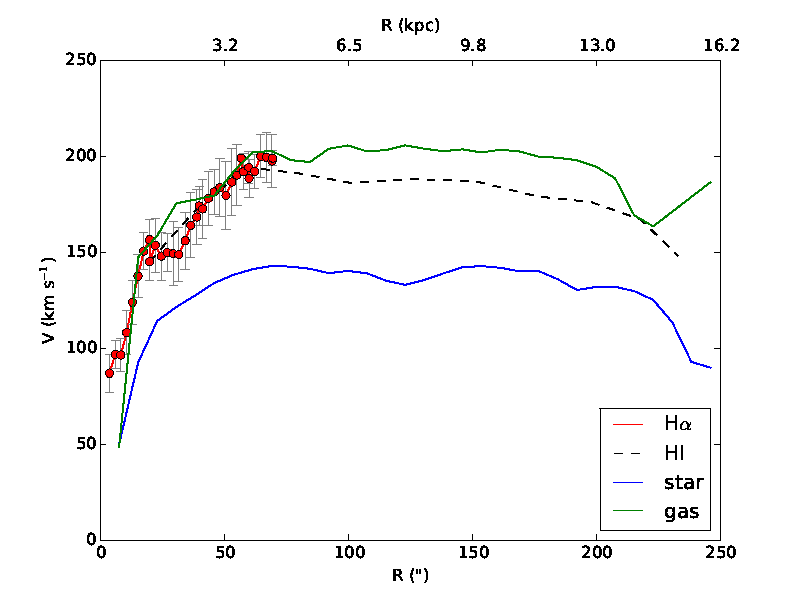}
\figcaption{Final rotation curve of the stellar (blue) and gas (green) disks of a host 
galaxy with a rigid halo compared to the observed H$\alpha$ (red dots) and HI (dashed)
 rotation curves.
\label{rcrig}}
\end{figure}

The matched position was found at $1.64$ Gyrs and is shown in Fig.\ref{momapsrig}. We had to rotate 
the image clockwise and adjust the origin.
 Ascan be noted, the morphology of the host is now quite different, showing no internal
 stellar spiral arms, but a very symmetric and open two armed gas spiral in the center.
 The tidal debris
 also appeared along the satellite passage track (north-west and south-east) and the tidal arms 
 are now more diffuse. When seen edge-on the stellar disk appears to be very thick.
The velocity field of the stellar disk is more irregular, and some additional smoothing was applied,
whereas the gas velocity contours resemble a more even shape.
 The major difference is observed in the shapes of the RCs in Fig.~\ref{rcrig}, where 
 the whole stellar RC is decreased by $\sim 60$ km s$^{-1}$. We did not explore the
 reason for this decay but speculate that the angular momentum of circular motion
 was transferred into the random motion of stellar particles. Despite this discrepancy, the gas RC shows
a very good match with the observed one.

\section{Discussion}
Simulations of the interaction process by self-consistent models remain 
a non-trivial task despite the growing computing power. In particular, reproducing observed
interacting galaxies is complicated beacause of the absence of full coordinates in phase space, and 
indirect methods have to be used to infer them.
The main difficulty arises beacause of a huge parameter space of initial conditions and 
the possibility of multiple solutions.
Recently, several advances were made in this direction that allowed faster exploration 
of the orbital parameters \citep{Petsch2008,Barnes2009}. A suitable solution can be rapidly found
if some restrictions are imposed. In particular, in this work we assume that the pair is
isolated and undergoes the first close approach following a nearly parabolic orbit;
the grand design spiral arms are trailing and were formed recently due to interaction.
Furthermore, we set the host to satellite mass ratio found from simulations with a simpler setup
as minimum necessary to induce the spiral pattern formation.

In this work we followed the iterative process of searching for orbital parameters, and 
with the imposed restrictions, a fast convergence was reached.
The single passage scenario of interaction is plausible under the assumptions above.
 We were able to reproduce the projected separation and systemic velocity difference between the
members of  KPG 302. In addition, we reproduce some faint features, such as the tidal eastern 
spiral arm and bridge between the galaxies. We also predict the existence of 
a faint spiral arm on the west side of the host disk, some caustics on the east side, and
a huge tail extending south-east toward nearby galaxy NGC 3906.
The velocity field of the host is very regular and symmetric inside the inner $5$ kpc disk.
 We also calculated the values for the pattern speed and corotation radius which are
  in marginal agreement with the previous numerical study of \citet{Kranz2003a}.

However, several differences remain concerning the shape and the extent
of the spiral arm of the host, and also its RC. We found that our model 
produces tightly wound arms with the pitch angle $\kappa=8\degr$, two times smaller than
the observed one. A detailed investigation of disk dynamics, resonances analysis, and
the influence halo properties is necessary to determine whether this is an 
intrinsic property of the galaxy model or the result of a particular interaction configuration.

The major discrepancy of our model comes from the mismatch of the kinematic major axis of the satellite.
 After the first passage the satellite disk changes orientation, tending to align 
its PA with the host PA. We tried to vary
the initial inclination and PA of the disk, to change the direction of rotation 
but were unable to reproduce correctly the velocity field.
We guess that the reason for that is a very massive satellite halo that transfers the 
intrinsic angular momentum acquired from the orbital motion to the disk particles.

The validity of a single passage scenario may be seriously questioned if we keep in mind the 
fact that the dynamical mass ratio of the pair is close to $100$ \citep{Fuentes2007}.
Our previous models where the companion was a point mass showed that in order to 
induce the spiral arm of the host the companion mass should be an order of magnitude 
bigger than the dynamical mass estimated from the RC \citep{Rosado2011}.
Further tests with mass ratios of $0.025$ and extended companions were unable to 
induce the spiral pattern in a single passage, indicating the necessity for multiple passages.
The fiducial collision model presented in this work was evolved until the satellite was 
completely merged with the host at $t=1.5$ Gyr, indicating that such mass ratio ($1:5.4$) 
and impact parameter ($5.6$ kpc) lead to a fast merging process and leaves no time for 
multiple passage.

On the other hand, the possibility of multiple passages may fail since the assumption of isolated
 evolution may not longer be valid since the pair is located a in relatively dense environment and 
 the neighbor galaxies may have been involved. Some indications of tidal interactions, such 
 as disk asymmetry may be homogenized and vanish in a dynamical time.
\citet{Pedrosa2008} showed through numerical simulations that the encounters produce bifurcated 
RCs if measured on opposite sides of the disks and that their symmetry is restored 
for $t\geq 1$ Gyr after the encounter.
Taking into account that our simulation took less than $1$ Gyr to reach the current
configuration and given that the observed RC shows a small asymmetry that is probably caused by interaction, 
we conclude that the scenario of a single passage is reliable. We, however, do not
exclude a scenario of spiral arm formation by a lighter satellite revolving on some 
resonance elliptic orbit.

The resemblance of M\,51-type systems is similar in many aspects and the obtained results are 
quite generic and may be compared to other systems. For example, KPG 302 is very similar
in appearance to NGC 7603 (ARP 92). We plan to continue this study by including more
realistic models and implementing optimization procedures that will reduce the 
computational cost to determine orbital parameters for other interacting pairs.

\acknowledgments We thank an anonymous referee for helpfull suggestions
on additional models and comments that improved the paper.
R.F.Gabbasov is supported by postdoctoral
fellowship provided by CONACyT. M.Rosado acknowledges CONACyT, project
number 46054-F, and DGAPA, project number IN 100606. The
simulations were performed on KanBalam cluster at DGSCA,
departament of supercomputing of UNAM and Atocatl cluster at 
Instituto de Astronomia UNAM. This work was partially supported by ABACUS, 
CONACyT grant EDOMEX-2011-C01-165873.

\nocite{*}
\bibliographystyle{apj}
\bibliography{bib}
\end{document}